*Article*

# The Complex Structure of the Abell 548–Abell 3367 Region

Mark J. Henriksen * and Layla Ahmed

Physics Department, University of Maryland, Baltimore County, MD 21250, USA; laylaiman05@gmail.com
* Correspondence: henrikse@umbc.edu

**Abstract:** Archival XMM and ROSAT X-Ray data are used to investigate the structure of the Abell 548–Abell 3367 region. Based on previous optical studies, this is a region likely to be rich in structure, although studies are in disagreement regarding the connection between Abell 3367 and Abell 548. We use the available archival X-Ray data together with kinematic data of counterpart galaxies to address this question and determine the structure in this region. The region is particularly rich in X-Ray structure elongated along a SW-NE axis and consisting of numerous extended X-Ray sources. In general, the structure consists of many galaxy groups and clusters which appear segregated in X-Ray luminosity, with the least luminous ~30% toward the outer region of the clusters, possibly tracing a filament. We find evidence to suggest a supercluster of three clusters at redshifts ~0.04, 0.045, and 0.06. Several of the X-Ray sources coincident with Abell 3367 have counterpart galaxy redshifts consistent with Abell 548, while others are significantly higher. This supports the formation of a supercluster consisting of Abell 548 and Abell 3667, with the higher-redshift X-Ray source being a background object. In addition, they are part of a larger structure consisting of a previously identified cluster at redshift 0.04 and two groups at redshift ~0.06. There is also filamentary structure at z ~0.103. The ubiquity of groups in the large-scale structure suggests that they provide an environment where galaxies are in close proximity and evolution via interaction can proceed well before the galaxies make their way into the dense central region of a cluster.

## 1. Introduction

Numerical simulations suggest that, in a Λ-CDM universe, a large fraction (up to 50%) of the baryonic matter remains located in a hot (0.01 to 1 keV) and tenuous ($\Omega_b \simeq 0.0068$) gaseous phase in the intercluster medium [1]. According to these same simulations, this gas is not uniformly dispersed between the clusters of galaxies, but is instead located in web-like structure composed of filaments, along with the dark matter and a significant fraction (~25%) of the galaxy population. Consequently, as inferred from primeval nucleosynthesis arguments, a significant fraction of the baryon content of the universe remains poorly characterized observationally.

Clusters are connected of filaments that consist of gas and galaxies, which are part of a complex dynamic accretion mechanism involving material falling into clusters [2]. Galaxies in filaments have been shown to be aligned with the principal axis of the distribution of galaxies in both clusters [3] and galaxy groups [4]. Groups have also been found to be preferentially aligned with the nearest Abell clusters. These alignments suggest that galactic evolution begins in the dynamic environment associated with filamentary structures.

New methods [5] and algorithms [6] are being developed to characterize the filamentary Universe. Because infalling gas from filaments has lower entropy than the rest of the cluster at the same radius, we should detect low-entropy fingers from the filaments entering the cluster outskirts. In fact, filament detection has been reported in the X-Ray [7,8], optical [9], and microwave [10] bands.

The outskirts of galaxy clusters and their smooth connections to the filamentary universe is a topic of considerable current interest. Because of the depth of the cluster gravitational potential well, the outer regions of clusters are heated to X-Ray emitting temperatures through accretion shocks. Both stellar and hot gas are probed in clusters and filaments in an effort to make precise measurements of the amount and distribution of the baryons. Components include hot gas, the stellar component in galaxies, and halo stars in lower halo masses [11]. A recent direction involves hot baryons on galactic scales, where hot and cool phases both exist in a circumgalactic halo [12].

Abell 548 and Abell 3367 may form a supercluster [13,14] pair separated by only 76 arcmin in projection, or 3.9 Mpc (at z = 0.0416 and $H_0$ = 70 km s$^{-1}$ Mpc$^{-1}$). Because of their proximity, they have been conjectured to be an interacting pair. Using 319 galaxy redshifts, [15] identified three main velocity components in the supercluster region: (1) Abell 3367 ($<v>$ = 30,477 km s$^{-1}$); (2) Abell 548 (double-peaked, $<v>$ = 11,951 km s$^{-1}$, and $<v>$ = 13,498 kms$^{-1}$ with a combined $<v>$ = 12,866 km s$^{-1}$; and (3) an unidentified filament region cluster ($<v>$ = 41,603 kms$^{-1}$). These authors concluded that the separation in velocity of 17,611 km s$^{-1}$, between Abell 548 and Abell 3367 indicates that the system is not close or interacting. However, this conclusion is at odds with [13,16], who placed Abell 3367 within 1000 km s$^{-1}$ of Abell 548.

The galaxies of the Abell 548–Abell 3367 region display a SW–NE pattern increasing in both RA and Dec (See Figure 1); therefore, this region represents an excellent candidate to contain a filamentary structure. This pattern is also evident in the X-Ray emission (See Figure 2), which is a powerful tool in mapping the structure in this region. There is no complete X-Ray coverage of this proposed supercluster other than that obtained with the ROSAT All-Sky Survey (RASS), which is shown in Figure 2.

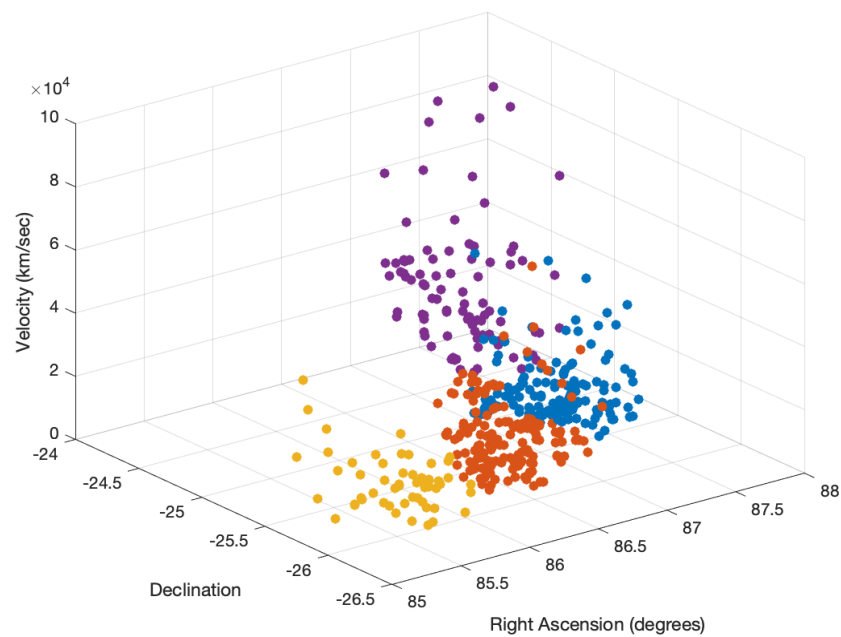

**Figure 1.** There are 481 galaxies detected with WISEA that have catalogued redshifts in the examined regions of Abell 548–Abell 3367. The colors correspond to the regions analyzed in the X-Ray band: Abell 548-A (blue), Abell 548-B (yellow), Abell 548-C (orange), and Abell 3367 (purple).

In this paper, we analyze archival X-Ray and optical–infrared data to better determine the substructure in this complex region.

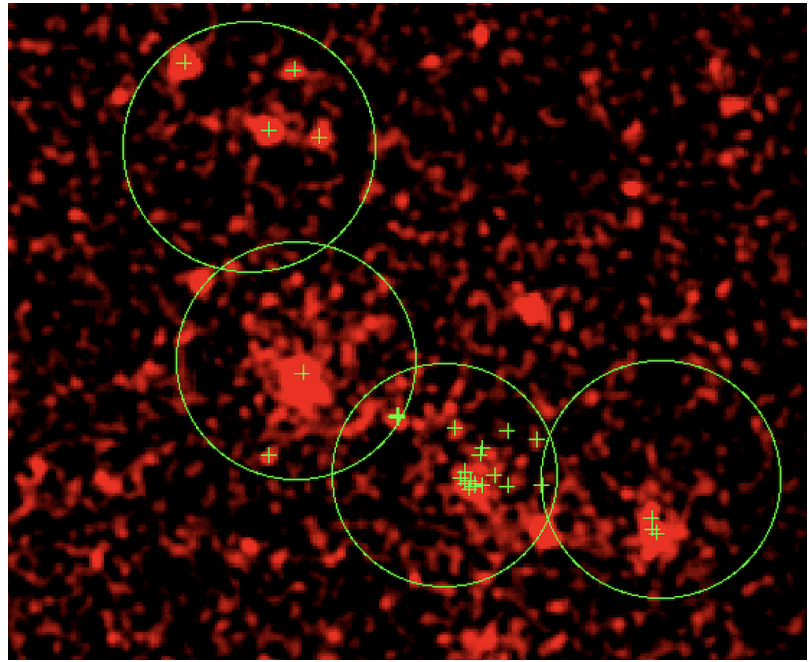

**Figure 2.** Extended sources (green crosses) overlaid on the smoothed RASS of the Abell 548–Abell 3367 region. The large circular regions are those used to extract sources from the XMM, ROSAT pointed, and RASS catalogs.

## 2. X-Ray Data

We used catalogued XMM serendipitous sources and ROSAT sources from both pointed and survey observations, as this is the existing X-Ray coverage of the Abell 548–Abell 3367 region. Four circular regions with 30 arcmin radius (1.5 Mpc at the distance of Abell 548) were used to characterize the region's X-Ray emission. We refer to these four regions as Abell 548-A, -B, and -C and Abell 3367. Two of these extraction regions, Abell 548-B and Abell 548-C, are covered by the 4XMM DR-12 Catalog [17]. These two extraction regions are centered at (5:46:0.0, −25:53:20.0) for Abell 548-B and at (5:42:0.0, −25:53:59.0) for Abell 548-C. Count rates were obtained with the EPIC detector for the 0.2–12 keV energy band for extended sources. There are 19 extended X-Ray sources in the Abell 548-B and -C regions. The raw data on RA and Dec in J2000 coordinates, extent in arc sec, count rate, and count rate error are provided in Table 1. The count rates were converted to flux assuming an APEC thermal model using the PIMMStool (https://heasarc.gsfc.nasa.gov/docs/software/tools/pimms.html, accessed on 1 November 2024), which is part of the HEASARC located at NASA's Goddard Space Flight Center. The APEC model used in the conversion has the following nominal parameters: kT = 1 keV, solar abundances, column density = $1.58 \times 10^{20}$ [18], and z = 0.0416 [19]. The same energy band was used for the count rate and flux. The flux was then converted to luminosity using the luminosity distance of 185.0 Mpc for $H_0$ = 69.6, $\Omega_M$ = 0.286, $\Omega_{vac}$ = 0.714 [20]; the angular scale at that distance is 0.827 kpc arcsec$^{-1}$. Also provided in Table 1 are the following derived parameters: extent in kpc, signal-to-noise, flux, and luminosity.

The region around Abell 548-A (5:48:46.23, −25:24:15.32) was searched for sources using the 2000 version of WGACAT [21], which uses PSPC pointed observations, and the 2RXS catalogue of point sources [22] from the RASS. Both energy bands are 0.1–2.4 keV. The raw data and calculated parameters for the eight sources are provided in Table 2 as for the Abell 548-B and -C regions in Table 1. However, these data are without quantitative extent measurement. The derived parameters were converted as described above using PIMMS. The point spread function of the pointed ROSAT PSPC/telescope system ranges from approximately 25 arcsec on-axis to 2 arcmin at 30 arcmin off-axis. The WGACAT point sources identified as "galaxy cluster" in the Abell 548-A region are all in the range of 30–60 arcmin from the center of the field as that is the only overlapping region with pointed observations. Although these are point sources, their size is unconstrained on scales less than 100 kpc. Based on averaging the sizes provided in [23], the characteristic optical diameter for compact groups is 181 ± 75 kpc. This is comparable to the resolution of the sources identified as point sources, meaning that a group of galaxies could be classified as a point source in this catalog based only on spatial information. The RASS has a PSF of 1.8 arcmin or 90 kpc, and as such has a similar ambiguity regarding extended versus true point sources. The luminosity is inconsistently high at $10^{37}$–$10^{39}$ ergs s$^{-1}$ [24] compared with bright point sources such as X-Ray binaries, and we evaluate them here as extended sources.

**Table 1.** Abell 548-B and -C extended X-Ray sources detected with XMM.

| No | RA | DEC | Extent | Extent | Count Rate | Error | SNR | Flux | $L_x$ | Log($L_x$) |
|---|---|---|---|---|---|---|---|---|---|---|
| | H:M:S | D:M:S | arc sec | kpc | $10^{-2}$ cps | $10^{-3}$ cps | | $10^{-14}$ ergs cm$^{-2}$ s$^{-1}$ | $10^{41}$ ergs s$^{-1}$ | ergs s$^{-1}$ |
| x1 | 5:44:51.2 | −25:41:56.0 | 6.77 | 5.60 | 1.11 | 2.2 | 5.09 | 1.27 | 5.10 | 40.71 |
| x2 | 5:45:19.0 | −25:46:09.9 | 19.4 | 16.1 | 1.89 | 1.81 | 10.5 | 2.17 | 0.87 | 40.9 |
| x3 | 5:45:19.0 | −25:55:35.6 | 80.0 | 66.2 | 4.95 | 13.0 | 3.80 | 5.66 | 2.28 | 41.36 |
| x4 | 5:45:22.0 | −25:47:51.0 | 18.8 | 15.5 | 0.86 | 1.94 | 4.39 | 0.98 | 0.39 | 40.59 |
| x5 | 5:45:27.6 | −25:55:16.2 | 80.0 | 66.2 | 3.69 | 18.0 | 2.05 | 4.22 | 1.70 | 41.23 |
| x6 | 5:45:28.4 | −25:55:48.2 | 37.2 | 30.8 | 22.8 | 11.3 | 20.2 | 26.10 | 10.50 | 42.02 |
| x7 | 5:45:32.7 | −25:54:23.2 | 49.5 | 40.9 | 4.27 | 7.96 | 5.36 | 4.88 | 1.97 | 41.29 |
| x8 | 5:45:33.3 | −25:56:28.6 | 58.8 | 48.6 | 24.8 | 15.8 | 15.7 | 28.30 | 11.40 | 42.06 |
| x9 | 5:45:38.5 | −25:52:13.3 | 15.8 | 13.0 | 0.59 | 1.55 | 3.81 | 0.67 | 0.27 | 40.43 |
| x10 | 5:45:38.6 | −25:55:05.4 | 27.2 | 22.5 | 1.65 | 3.32 | 4.95 | 1.88 | 0.76 | 40.88 |
| x11 | 5:45:43.6 | −25:53:44.8 | 80.0 | 66.2 | 4.82 | 18.4 | 2.63 | 5.52 | 2.22 | 41.35 |
| x12 | 5:45:49.8 | −25:41:08.6 | 44.2 | 36.5 | 6.14 | 13.3 | 4.63 | 7.02 | 2.83 | 41.45 |
| x13 | 5:45:05.6 | −25:53:08.8 | 14.1 | 11.6 | 0.63 | 1.24 | 5.06 | 0.84 | 0.34 | 40.53 |
| x14 | 5:44:51.5 | −25:55:44.9 | 11.2 | 9.27 | 0.77 | 1.00 | 7.67 | 1.03 | 0.41 | 40.62 |
| x15 | 5:44:12.6 | −25:55:34.7 | 22.5 | 18.6 | 13.4 | 2.67 | 5.02 | 17.60 | 6.80 | 41.83 |
| x16 | 5:44:18.4 | −25:43:55.4 | 10.6 | 8.75 | 0.40 | 1.73 | 2.3 | 0.53 | 0.21 | 40.33 |
| x17 | 5:42:10.0 | −26:03:37.2 | 13.4 | 11.1 | 2.03 | 4.00 | 5.08 | 2.72 | 1.10 | 41.04 |
| x18 | 5:42:09.3 | −26:06:23.4 | 30.0 | 24.8 | 3.57 | 12.6 | 2.83 | 4.77 | 1.92 | 41.28 |
| x19 | 5:42:04.28 | −26:07:27.5 | 50.8 | 42.0 | 44.9 | 17.1 | 26.3 | 60.2 | 24.3 | 42.38 |

**Table 2.** Abell 548-A extended X-Ray sources detected with ROSAT Pointed and RASS.

| No | RA | DEC | Count Rate | Error | SNR | Flux | $L_x$ | Log($L_x$) | Extent |
|---|---|---|---|---|---|---|---|---|---|
| | H:M:S | D:M:S | $10^{-2}$ cps | $10^{-3}$ cps | | $10^{-14}$ ergs cm$^{-2}$ s$^{-1}$ | $10^{41}$ ergs s$^{-1}$ | ergs s$^{-1}$ | kpc |
| x20 | 5:47:10.8 | −25:35:20 | 0.57 | - | - | 5.53 | 2.23 | 41.35 | cluster |
| x21 | 5:46:55.0 | −25:38:10 | 1.77 | - | - | 17.2 | 6.92 | 41.84 | cluster |
| x22 | 5:46:55.2 | −25:38:16 | 1.27 | - | - | 12.3 | 4.97 | 41.70 | cluster |
| x23 | 5:46:54.1 | −25:38:02 | 2.26 | - | - | 21.9 | 8.84 | 41.95 | cluster |
| x24 | 5:46:53.3 | −25:38:40 | 2.47 | - | - | 24.0 | 9.66 | 41.98 | cluster |
| x25 | 5:48:38.91 | −25:27:21.4 | 29.4 | 26.8 | 11.0 | 286. | 115. | 43.06 | - |
| x26 | 5:49:17.07 | −25:47:59.7 | 3.71 | 9.9 | 3.8 | 36.0 | 14.5 | 42.16 | - |
| x27 | 5:46:52.90 | −25:38:29.4 | 4.96 | 1.23 | 4.0 | 48.1 | 19.4 | 42.29 | - |

Abell 3367 is at z = 0.1016. For the adopted cosmology, the luminosity distance is 471.4 Mpc and the angular scale is 1.88 kpc arc s$^{-1}$. RASS provides the only available coverage of Abell 3367. A 30 arcmin radius circle centered at (5:49:37.14, −24:30:0.0) was searched for sources using two catalogs: the 2RXS catalog of point sources, and the RASS faint source catalog [25]. The raw data and derived parameters for the five sources are provided in Table 3 for the Abell 548-B and -C regions in Table 1, without indication of extent. At the distance of Abell 3367, a point source is 200 kpc in radius. Thus, a point source could be an unresolved extended source, such as a galaxy group with characteristic optical diameter of 181 ± 75 kpc [23]. The range of luminosities also supports this interpretation.

Table 3. Abell 3367 extended X-Ray sources detected with RASS.

| No | RA<br>H:M:S | DEC<br>D:M:S | Count Rate<br>$10^{-2}$ cps | Error<br>$10^{-3}$ cps | SNR | Flux<br>$10^{-14}$ ergs cm$^{-2}$ s$^{-1}$ | $L_x$<br>$10^{41}$ ergs s$^{-1}$ | Log($L_x$)<br>ergs s$^{-1}$ |
|---|---|---|---|---|---|---|---|---|
| x28 | 5:48:47.6 | −24:10:42.3 | 3.58 | 10.9 | 3.3 | 29.9 | 82.7 | 42.92 |
| x29 | 5:48:19.7 | −24:27:46.4 | 7.07 | 13.7 | 5.2 | 59.0 | 163 | 43.21 |
| x30 | 5:50:47.4 | −24:08:43.2 | 39.6 | 28.7 | 13.8 | 331 | 915 | 43.96 |
| x31 | 5:49:15.4 | −24:25:50.9 | 29.8 | 2.55 | 11.7 | 249 | 689 | 43.84 |
| x32 | 5:49:07.9 | −24:20:24.0 | 1.41 | 6.86 | 2.1 | 11.8 | 32.6 | 42.51 |

## 3. Identification of Counterparts to Extended X-Ray Sources

There are 481 WISEA [26] galaxies with redshifts in the four regions we examined in the X-Ray band. WISE has mapped the sky in four infrared bands: 3.4, 4.6, 12, and 22 µm. The galaxies exhibit a dense and continuous distribution in 3D (see Figure 1).

Counterparts that are spatially coincident with the extended XMM sources were identified to aid in characterizing the X-Ray emission (e.g., galaxy, group, or subcluster). The X-Ray luminosity was also used for identification (see Section 4). The counterparts were found in the NASA Extragalatic Database. Spatial coincidence occurs when the coordinates of a counterpart fall within the boundaries of an X-Ray sources using the position and size of each source (see Table 4). Complete size information is available for XMM; however, for ROSAT the catalogs either have no available information on the extent or are characterized as point sources. In the latter case, size estimates were based on the point-spread function. Identifications were also aided by the similarity of the X-Ray luminosity with a particular class of object.

Table 4. Number of counterparts for extended X-Ray sources.

| No | Extent<br>arc sec | Extent<br>kpc | No. Galaxies | Mean Redshift | Error |
|---|---|---|---|---|---|
| x1 | 6.77 | 5.60 | - | - | - |
| x2 | 19.4 | 16.1 | 1 | - | - |
| x3 | 80.0 | 66.2 | 23 | 0.0461 | |
| x4 | 18.8 | 15.5 | 2 | - | - |
| x5 | 80.0 | 66.2 | 37 | 0.044 | 0.001 |
| x6 | 37.2 | 30.8 | 11 | 0.043 | 0.001 |
| x7 | 49.5 | 40.9 | 14 | - | - |
| x8 | 58.8 | 48.6 | 23 | 0.0448 | 0.0005 |
| x9 | 15.8 | 13.0 | 2 | - | - |
| x10 | 27.2 | 22.5 | 3 | - | - |
| x11 | 80.0 | 66.2 | 31 | 0.058 | 0.022 |
| x12 | 44.2 | 36.5 | 13 | 0.1029 | - |
| x13 | 14.1 | 11.6 | - | - | - |
| x14 | 11.2 | 9.27 | 1 | 0.0487 | - |
| x15 | 22.5 | 18.6 | 5 | 0.26 galaxy cluster? | |
| x16 | 10.6 | 8.75 | 2 | - | - |
| x17 | 13.4 | 11.1 | - | - | - |
| x18 | 30.0 | 24.8 | - | - | - |
| x19 | 50.8 | 42.0 | 2 | 0.043 | 0.006 |
| x20 | - | cluster | 4 | 0.06175 | 0.0386 |
| x21 | - | cluster | 3 | 0.0454 | 0.0013 |
| x22 | - | cluster | 3 | 0.0454 | 0.0013 |
| x23 | - | cluster | 3 | 0.0454 | 0.0013 |

**Table 4.** *Cont.*

| No | Extent arc sec | Extent kpc | No. Galaxies | Mean Redshift | Error |
|---|---|---|---|---|---|
| x24 | - | cluster | 3 | 0.0454 | 0.0013 |
| x25 | - | - | 7 | 0.0463 | 0.0179 |
| x26 | - | - | 0 | - | - |
| x27 | - | - | 3 | 0.0454 | 0.0013 |
| x28 | - | - | 2 | 0.1029 | - |
| x29 | - | - | 1 | - | - |
| x30 | - | - | 1 | 0.0442 | - |
| x31 | - | - | 1 | 0.0450 | 1 |
| x32 | - | - | 1 | 0.2673 | - |

## 4. Results

Figure 2 displays the extended X-Ray source locations (crosses) withing the four circular regions examined for extended X-Ray emission overlaid on the smoothed RASS image. The RASS image is centered on (5:45:38.56, −25:34:42.57). The image size is 2.3 × 1.9 degrees. Extended sources stretch across the region co-spatially with visible clumps of hot gas, highlighting fainter regions of emission.

The histograms in Figures 3–5 show the X-Ray sources from Abell 548-A, -B, and -C from left to right. Abell 548-A has few sources below $\log(L_x) < 41.5$, while Abell 548-B has few above 41.5. Abell 548-C has two below and one above.

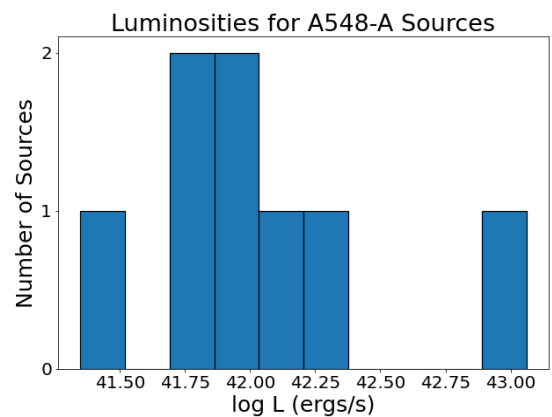

**Figure 3.** Abell 548-A sources.

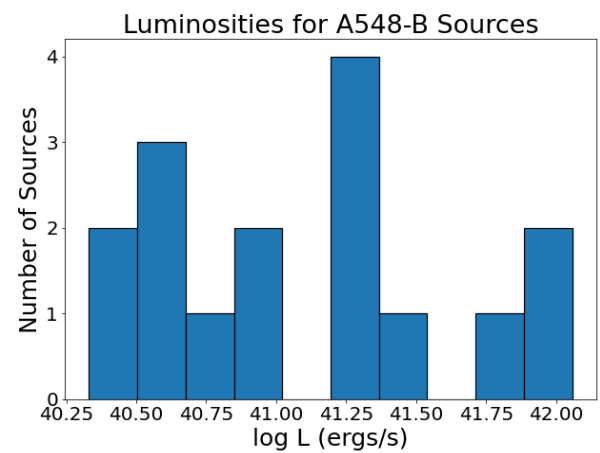

**Figure 4.** Abell 548-B sources.

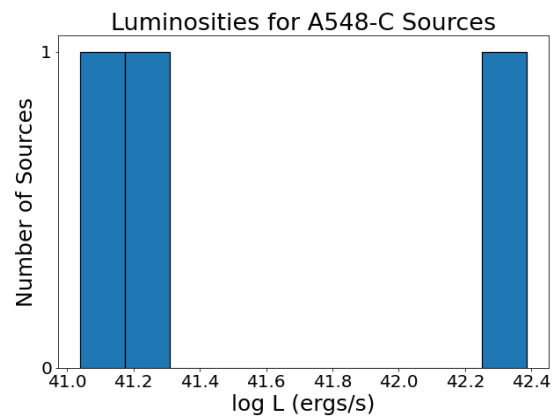

**Figure 5.** Abell 548-C sources.

Figure 6 shows the luminosity histogram of the sources in the Abell 3367 region. The majority are above log($L_x$) = 43 and significantly higher than the Abell 548 sources. The data consist of faint RASS sources and RASS point sources. The point source resolution was 200 kpc at the redshift of Abell 3367, which implies that the sources could be as large as groups. The higher luminosity could be due to the fact that the data consist entirely of short RASS exposures, meaning that that only the brightest sources are detected. Thus, the brightest sources, for example subclusters, are favored for detection. On the other hand, they could be true point sources such as AGNs. AGNs located outside of the cluster center are unlikely; however, recent data show that this occurs in high-Z clusters with Z > 1 [27]. With high-quality spectra, the type of source could be determined.

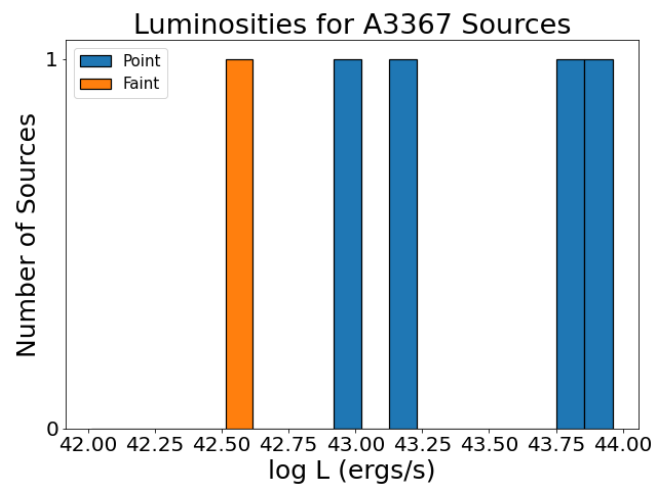

**Figure 6.** RASS detections from the point source and faint source catalogs. The PSF is 200 kpc at this redshift, meaning that that an extended source such as a galaxy group might appear as a point source.

Figure 7 shows the full range of the extended X-Ray source luminosities from ~40.5 to ~43.8, spanning four orders of magnitude. The color-coded regions clearly show the offset and overlap between regions; however, there is a visible concentration in the range of 40.5–41.5.

In addition to X-Ray luminosity and size, we searched for counterparts at other wavelengths following the procedure discussed in Section 3. The results are provided in Table 4. Using the NASA Extragalactic Database, 85% of the X-Ray sources were found to be overlapped by galaxies in the WISEA catalog, while 66% of the X-Ray sources had a mean redshift obtained from averaging counterpart galaxies with a reported redshift. Table 4 shows the number of overlapped galaxies and the calculated mean and standard deviation of the sources' counterpart redshifts. There are four X-Ray sources with a large number of redshifts: x3 and x8 with 23 galaxies each, x11 with 31 galaxies, and x5 with 37 galaxies. These are also the four X-Ray sources with the largest spatial extents.

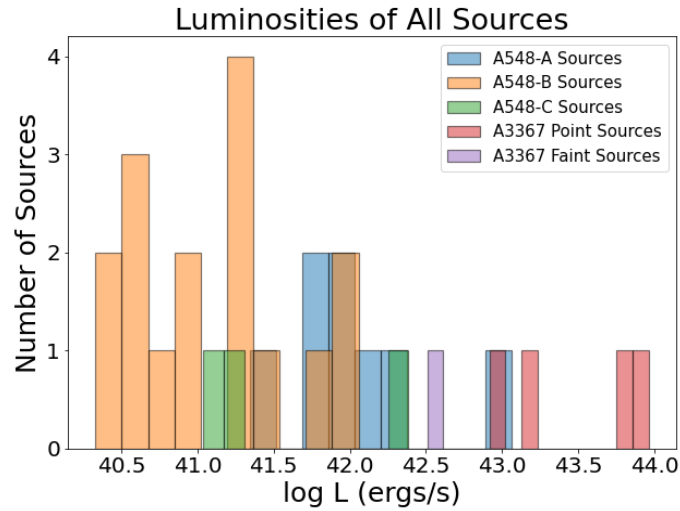

**Figure 7.** Luminosity distributions of all detected sources, showing a large spread in luminosity with a concentration at Log($L_x$) = 41.5.

Figure 8 compares the distribution of the X-Ray sources to a sample of galaxy groups [28]. These galaxy groups were chosen because they have a similar range of X-Ray luminosities compared to the extended X-Ray sources. The comparison sample consisted of approximately 75% detections of diffuse group X-Ray emissions and 25% upper limits distributed from Log($L_x$) ~40–42. Group X-Ray luminosities ranged from those comparable to elliptical galaxies on the low end to subclusters on the high end. The group X-Ray luminosity comparison sample peaked at 42.0. While the X-Ray subsample overlaps the group sample, in general the composite of all X-Ray sources shows a lower peak than the groups, at around 41.0.

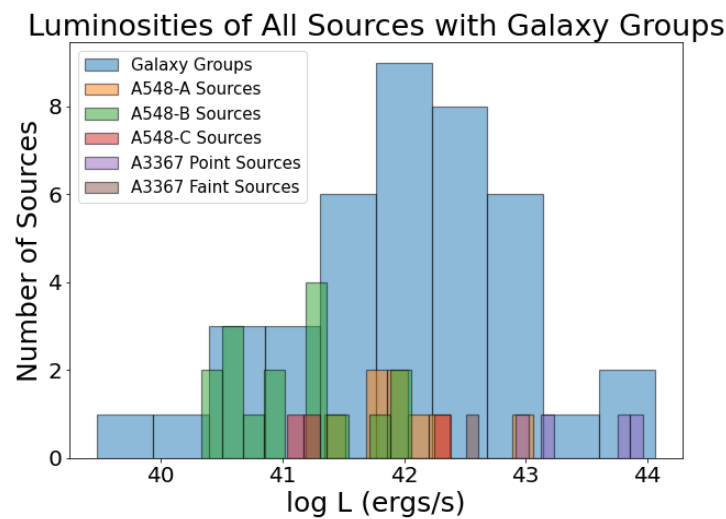

**Figure 8.** Comparison of all extended X-Ray source luminosities with groups of galaxies. The groups have a peak at Log($L_x$) = 42, slightly higher than the concentration of extended X-Ray sources but still reasonably consistent.

Figure 9 indicates the significant difference in the cluster regions from which the samples originate. The Abell 548-A sources shown in the right panel are close to the group mean, while the Abell 548-B sources in the left panel overlap the lower half of the group distribution. Spatially, the location of the galaxies for Abell A548-A and -B are indicated in Figure 1 by blue and yellow, respectively, with blue near the densest part of the system and yellow toward the lower-density end. The difference in positions can also be seen in Figure 10.

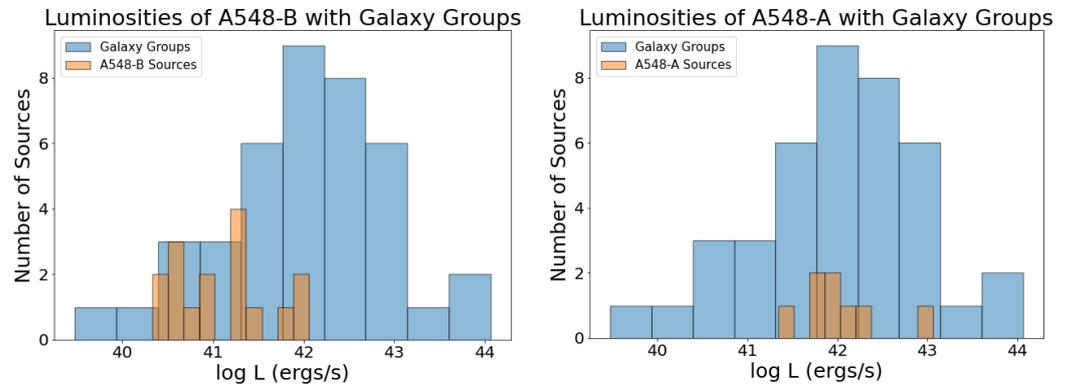

**Figure 9.** Abell 548-B (L) and Abell 548-A (R) sources compared to groups of galaxies. While the Abell 548-A sources match up well with the groups, the Abell 548-B sources are shifted significantly to lower X-Ray luminosity. The lower sources are preferentially found in the cluster outskirts.

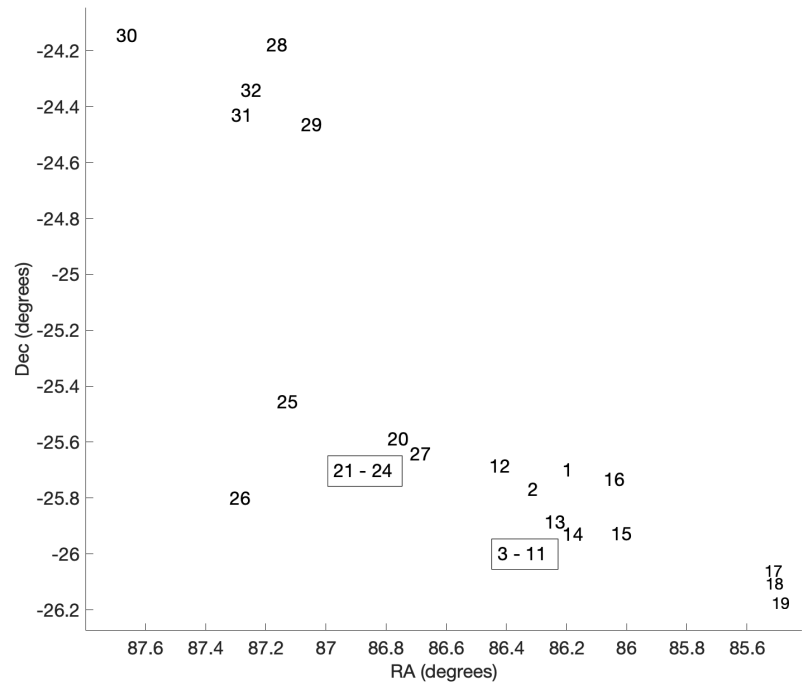

**Figure 10.** The location of X-Ray sources on the sky. Rectangles show the general area of sources plotted in adjoining panels. The locations on the sky are combined with redshift information to determine whether the sources are part of larger structures.

Figure 10 shows the spatial distribution of the X-Ray sources in the sky. The lowest-luminosity X-Ray sources are x4, x9, x10, x13, x14, and x16, which are the sources with luminosity less than $10^{41}$ ergs s$^{-1}$. These sources are located between 1.3–1.9 Mpc from the optical center of Abell 548, (86.749, −25.598) [29]. Thus, the difference in X-Ray luminosity may reflect that more massive and higher-luminosity galaxies tend to be found near the center of a cluster compared to the outskirts, that is, the well established morphology–density relationship that is apparent in both observations [30] and simulations [31]. The rectangles in Figure 10 contain closely grouped sources that are plotted in Figure 11 because the scale of Figure 10 cannot accommodate them.

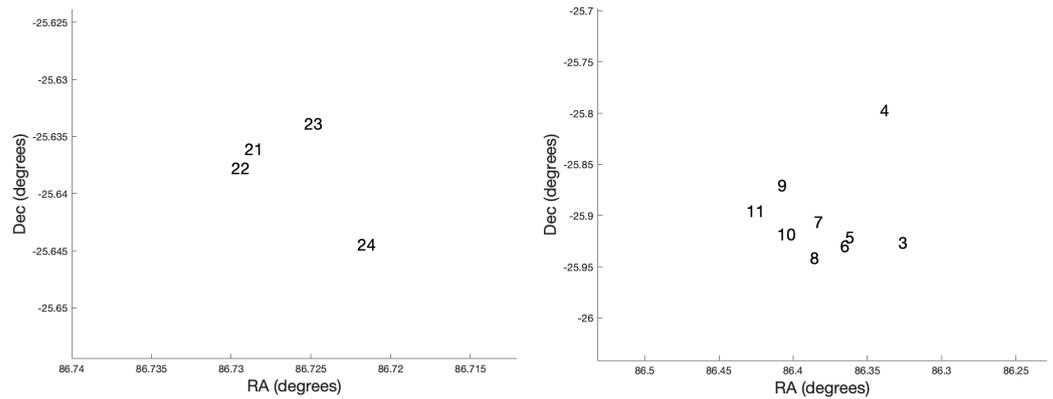

**Figure 11.** Left panel: sources 21–24; right panel: sources 3–11.

Figure 12 shows the Abell 548-C sources which bridge the Abell 548-A and -B sources in X-Ray luminosity. Of the three X-Ray sources, two are consistent with the rather low Abell 548-B luminosities and one is consistent with the higher Abell 548-A luminosities. All three are consistent with typical group luminosity. Figure 13 shows the Abell 3367 sources compared to the groups. These are the highest X-Ray luminosity sources in the region, and are consistent with the high end of the group luminosities.

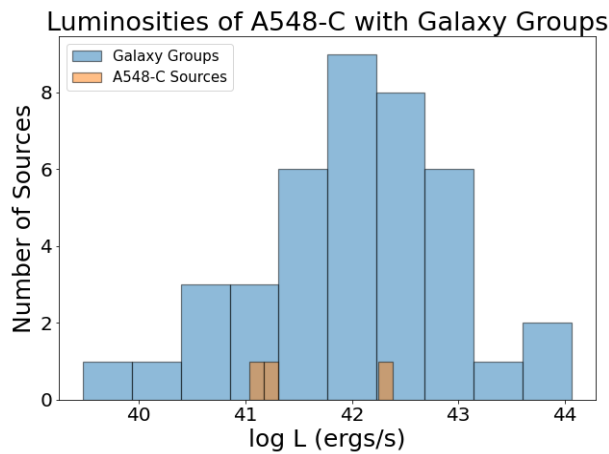

**Figure 12.** Comparison of Abell 548-C sources with groups of galaxies

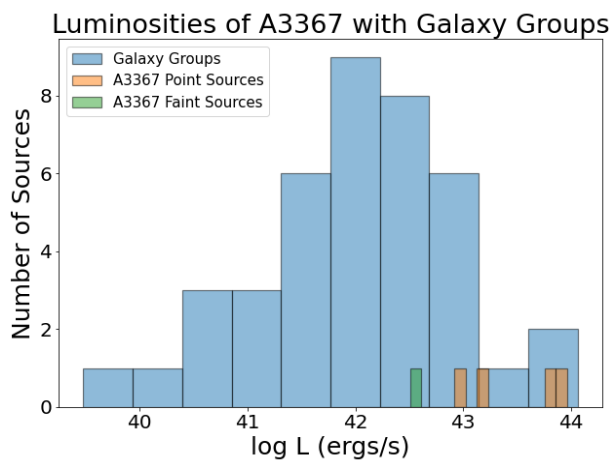

**Figure 13.** Comparison between Abell 3367 sources detected with RASS and groups of galaxies. The luminosities are towards the high end of the groups. Because the RASS observations are short, these high-luminosity sources are likely selected.

## 5. Discussion

The structure and dynamics of the putative supercluster region are not well established. In fact, the NASA Extragalactic Database has six wide-ranging redshifts for Abell 548 alone: 0.031, 0.042, 0.063, 0.087, 0.101, and 0.138. In the study of structure, X-Ray observations are valuable because galaxy structures on all scales contain trapped hot gas that emits X-Rays delineating the larger structures. Here, we use galaxies that are co-spatial with the X-Ray structures to provide redshift information for X-Ray sources. To investigate the larger redshift structure of the X-Ray sources, the counterpart galaxies are binned using the redshift. Figure 14 shows that the counterpart galaxies fall into four distinct redshift bins of 0.045, 0.06, 0.103, and 0.26 over the Abell 548–Abell 3367 region. An additional constraint for evaluating plausible large-scale structure is the separation of the X-Ray sources in the sky identified with these redshift structures. This is shown in Figures 10 and 11. There are two X-Ray sources (x20, and x11) in the Z ~0.06 bin, consisting of 4 and 31 counterpart galaxies, respectively. Their sky separation is 1.89 Mpc. The collection of 14 X-Ray sources (x3, x5, x6, x8, x14, x19, x21-25, x27, x30, x31) in the Z ~0.045 bin includes two X-Ray sources with high X-Ray luminosity (x30, x31) that are consistent with high luminosity groups and subclusters. The maximum separation of the X-Ray sources in this redshift bin is 2.89 Mpc. These 14 X-Ray sources have 121 counterpart galaxies. This structure has a galaxy-weighted mean redshift of 0.045 with a variance of 0.001, Thus there is no indication of significant redshift substructure within this redshift bin. These two lower-redshift bins cover a fairly large region, though they plausibly form a filamentary structure at a fairly low angle to the line of sight.

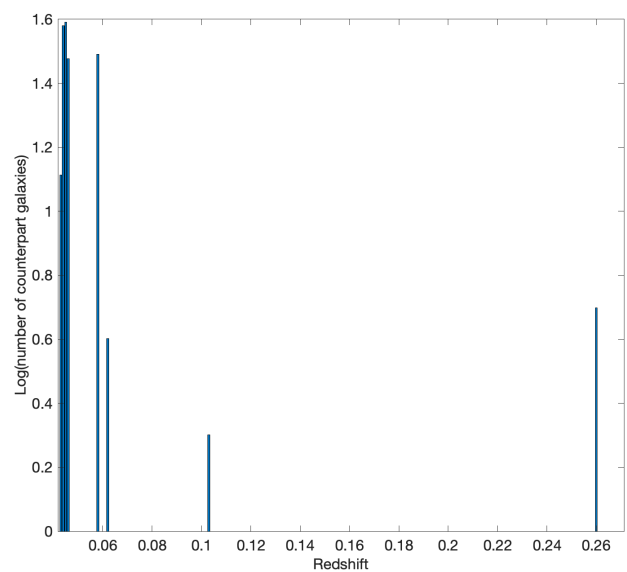

**Figure 14.** Histogram of counterpart galaxy redshifts. Four distinct structures in redshift are apparent at 0.045, 0.06, 0.103, and 0.26.

The separation of Z ~0.103 bin X-Ray sources in the sky is 11.3 Mpc. This structure consists of X-Ray source x12, with 13 counterpart galaxies, and x28, with two counterpart galaxies. Source x28 has a log ($L_x$) of 42.92, which is at the high end of the galaxy groups. Thus, they may trace a filamentary structure. Because there are only two sources in a relatively large region, deeper X-Ray observations of the intervening region are needed to verify whether filamentary gas is present. The highest redshift bin, 0.26, consists of X-Ray sources x32 and x15, with one and five counterpart galaxies, respectively. Their separation in the sky is 28.3 Mpc, which is quite large, and their redshift is quite disparate from the other structures in the region, making them likely to be two isolated groups in the region. It is unlikely that they are part of the suggested structures at Z ~0.045, 0.06, and 0.103.

Source x25 is co-spatial with a galaxy cluster at z = 0.040 [10] detected in the microwave, which is not identified with an Abell cluster. For $H_0 = 69.6$ km s$^{-1}$ Mpc$^{-1}$, $\Omega_m = 0.286$, $\Omega_{vac} = 0.714$, the luminosity distance is 177.6 Mpc. The Abell 548 cluster is at luminosity distance 185.0, suggesting that this cluster is part of the larger supercluster posited above. The six X-Ray sources in the Abell 3367 region all have optical counterpart galaxies with redshifts. Two of these galaxies are roughly consistent with the Abell 548 redshift, while two are much higher. We suggest that the Abell 3367 region has some connection to the Abell 548 cluster region along with some optical background contamination. Based on the X-Ray structure and counterpart redshifts, Abell 3367 is likely part of a larger system involving Abell 548 and the third cluster at z = 0.040.

Because there is a concentration of sources with luminosity below the group sample peak by roughly a factor of 10, we investigated other types of X-Ray sources as possible counterparts. An XMM survey of 198 galaxies [32] with mean redshift of 0.09 and composed of an equal mixture of E/S0 and Sa-Scd types provided a secondary comparison sample with lower X-Ray emission. For the galaxy sample, the mean log ($L_x$) for early-type (E/S0) and spiral (Sa-Scd) is 40.48 +0.12/−0.18 and 39.70 +0.25/−0.70, respectively. The mean of the Abell 548 sources is 41.41 +/− 0.70. Including the Abell 3367 sources, the mean is 41.72 +/− 0.97. The characteristic mean of the Abell 548–Abell 3367 X-Ray sources is a decade higher in luminosity than the mean of the brightest X-Ray E/S0 galaxies. Figure 15 shows the number of counterparts for a source versus its X-Ray luminosity. The horizontal line is at six counterpart galaxies, while the vertical line is at an X-Ray luminosity of $10^{41}$ ergs s$^{-1}$. It is apparent that most of the X-Ray sources with just a few galaxies all have low luminosities. In contrast, all of the X-Ray sources with more than five counterparts have higher X-Ray luminosity. This indicates that more galaxies are associated with higher X-Ray emission. This trend is consistent with X-Ray sources being associated with a galaxy group, though perhaps being dominated by a single elliptical galaxy at the lower-luminosity end, as in a fossil group [33]. Therefore, we conclude that the Abell 548–Abell 3367 supercluster sources are mainly galaxy groups.

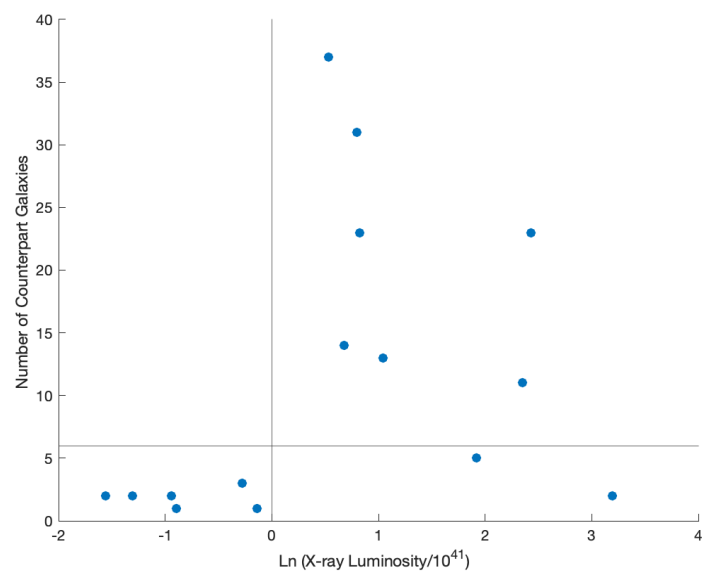

**Figure 15.** The horizontal line is at six counterpart galaxies. The vertical line is at X-Ray luminosity of $10^{41}$ ergs s$^{-1}$. Most of the X-Ray sources with just a few galaxies have low luminosity. All of the X-Ray sources with more than five counterparts have higher luminosity. This indicates that a higher number of counterparts is associated with higher X-Ray emission, consistent with the X-Ray source being associated with a group.

Although ~30% of the X-Ray sources are identified with low-luminosity galaxy groups, there are also several at the high end of the group luminosity function. The plethora of low-luminosity sources is likely due to their association with regions of low galaxy density, on the outskirts of the clusters, and along the filamentary region connecting the clusters. In this way, X-Ray emissions from galaxy groups and the baryons tied up within them trace the large-scale structure and not just the high-density regions of clusters.

For comparison, we cite another multiple-cluster filamentary system, Abell 3395–Abell 3391, which was discovered in the X-Ray using ROSAT pointed observations and the Advanced Satellite for Cosmology and Astrophysics [34]. Diffuse filamentary gas was positively detected in the X-Ray for this supercluster region. This system has a microwave detection of the intercluster region as well [10]. More recently, Chandra and XMM observations have confirmed the X-Ray filament and detected the group ESO-161, which lies on the filament [35]. eRosita, with increased soft X-Ray sensitivity, detected multiple supercluster components, including warm gas, clusters, and groups. Suzaku observations of the Abell 3395 cluster outskirts and the connecting filament had the aim of measuring the metal abundance [36]. The measured value was statistically consistent with the average value found for clusters in general, implying early enrichment in which galactic gas loss processes lead to gas enrichment while the cluster is still accreting material from the filament. The gas density is quite low, favoring galactic expulsion via winds over ram pressure stripping, although ram pressure has been reported in a group in the filament [35].

Massive galaxy clusters are late-forming structures, and their galaxies have a mean infall time of ~4.5 Gyr [37]. Of relatively recently infalling galaxies, 40% are in groups [38]. Star formation is also found to be suppressed within galaxy groups [39] that are later accreted into clusters. Evidence of this galaxy quenching is also seen in X-Ray point source studies of compact groups that show signs of galaxy interaction [40]. We have found that the region around Abell 548 and Abell 3367 is a complex amalgam of structure in the X-Ray band, marked by many galaxy groups with larger structures. Many of the groups are low-luminosity, implying low mass, likely due to their formation in low-density filamentary regions. The galaxies in these groups are evolving well before they enter the cluster. This is further highlighted by a fast radio burst that was recently identified within a compact galaxy group. This group shows signs of interaction among group members. The spectroscopic redshift of the system is z = 1.017 and the host galaxy for the fast radio burst is a star-forming galaxy [41]. We conclude that large-scale structure such as we have studied here is marked by many galaxy groups, which can lead to significant galaxy evolution before the galaxies reach the dense cluster environment.

**Author Contributions:** L.A. collected the data, analyzed it, and prepared the tables and figures. M.J.H. designed the project, supervised the analysis, and wrote the paper. All authors have read and agreed to the published version of the manuscript.

**Funding:** This research is not externally funded.

**Data Availability Statement:** There are no new raw data presented in this paper. Extracted and calculated parameters are accessible in the published tables.

**Conflicts of Interest:** The authors declare no conflicts of interest.